\documentclass[twocolumn,showpacs,amsmath,amssymb,floatfix,pre]{revtex4}
\usepackage{graphicx}
\usepackage{epstopdf} 

\begin{document}
\draft

\title{Relation between the High Density Phase and the Very-High Density
  Phase of Amorphous Solid Water}

\author{Nicolas Giovambattista$^1$, H. Eugene Stanley$^1$ and Francesco
Sciortino$^2$}

\affiliation{
$^1$Center for Polymer Studies and Department of Physics,
Boston University, Boston, MA 02215 USA \\
$^2$ Dipartimento di Fisica and INFM Udr and 
Center for Statistical Mechanics and Complexity,
Universita' di Roma ``La Sapienza'' \\
Piazzale Aldo Moro 2, I-00185, Roma, Italy
}

\date{11 March 2004}


\begin{abstract}
It has been suggested that high-density amorphous (HDA) ice is a
 structurally arrested form of high-density
 liquid (HDL) water, while low-density amorphous (LDA) ice 
is a structurally arrested form of
 low-density liquid (LDL) water.
Recent experiments and simulations have been interpreted to support
the possibility of a second ``distinct" high-density structural state,
named very high-density amorphous (VHDA) ice, questioning the LDL-HDL
hypothesis.  We test this interpretation using extensive computer
simulations, and find that VHDA is a more stable form of HDA and that in
fact VHDA should be considered as the amorphous ice of the
quenched HDL.
\end{abstract}

\maketitle



\bigskip\bigskip

\noindent
The most common form of water in the universe is not a liquid but is a
disordered solid named glassy water \cite{astro}.  Depending on the
glass formation route---vapor deposition, hyperquenching of the liquid,
pressure induced crystal amorphization---water forms amorphous solids of
quite different structure, with the density varying by as much as 40
percent between the lowest density form and the highest
\cite{angellrev,pablo}.  The structural and thermodynamical properties of
 the different amorphous forms of glassy water have been the focus of
 many recent experimental 
\cite{loerting2,finneyHDAVHDA,soperHDA100,science,loerting,%
finneyLDAHDA,mishimanature02,new}, numerical and
theoretical studies \cite{french,brazhkin,ourPRL}. 
However, a clear picture of the phase diagram of glassy water is still
missing. This is in part because the properties of glasses change
 drastically with aging\cite{science,mishimanature02} and because experiments
 at high pressure are difficult to perform.
Glassy water can be found in at least two different forms, low-density
amorphous (LDA) and high-density amorphous (HDA) ice
\cite{visual,mishimaNature85}. 
If a glass is formed by extremely rapid cooling of a liquid, then one
can naturally associate the glass with the liquid state.
LDA can be obtained experimentally by very fast quenching of
 low-density liquid (LDL) water
at normal pressure\cite{HGW}. However, the generation of an HDA glass via
high pressure cooling of the liquid have proved an elusive goal; in fact,
HDA is experimentally formed not by quenching but rather by compression
either of LDA or of crystalline ice \cite{mishimaNature85}. 
 Therefore, the actual relationship of glassy water to the liquid 
at high pressure remains unknown. 
On the other side, computer simulations \cite{poole,poolepre} indicate that HDA is a 
structurally arrested form of high-density liquid (HDL) water, and therefore, 
HDA can be obtained by quenching of HDL at high pressure.
A second ``distinct" high-density structural state,
named very high-density amorphous solid (VHDA) has been recently discovered.
The density of VHDA is 7-8\% higher than the
density of HDA.  VHDA is generated by heating HDA under pressure. The
resulting glass does not convert back to HDA when recovered at ambient
pressure at $T=77 $K.  The possibility of a VHDA phase in addition to an
HDA phase raises many interesting questions \cite{klug}, such as the
relation between VHDA and HDA; this relation is important to elucidate
the hypothesis that below a critical temperature there are two distinct
phases of liquid water, LDL and HDL \cite{poole}. 
 Which phase, HDA or VHDA, more resembles liquid
water at high pressure is particularly interesting, and it appears that
the structural properties of VHDA are closer to liquid water at high
pressure than those of HDA \cite{finneyHDAVHDA,french}. 
On the other side, the presence of two different high-density amorphous ices 
 could imply the existence of more than one form of HDL as recently 
 suggested by computer simulations \cite{geiger}.

Computer simulations offer a potentially useful tool to probe the
relation between HDA, VHDA and liquid water, since the time scale of
most simulations is sufficiently short that liquid and glassy states can
be studied for a wider range of state points than is accessible
experimentally.  Here we report a set of extensive simulations that
suggest that VHDA not HDA, may be considered as a physical manifestation of the
quenched high pressure liquid.  Further, our simulations suggest that
VHDA is not a new thermodynamically distinct structure but rather VHDA
results from partial annealing of the HDA structures made possible by
the higher annealing temperature.  Hence HDA is not stable but rather is
highly metastable, relaxing to VHDA in a fashion analogous to the way
that, on slow heating, glasses generated with hyperquenched methods
relax to glasses generated with standard cooling rates.

Simulations offer the unique possibility of comparing (i) the glass
resulting from conversion of HDA to VHDA with (ii) the glass generated
by isobaric cooling of the liquid. However, a note of caution is in
order, since in comparing experiments and simulations for glasses, one
must carefully account for the significantly different time scales
probed.  State-of-the-art simulations probe time scales of 10--100 ns,
so when the characteristic relaxation time becomes comparable to this
time, the system glassifies.  Since the experimental homogeneous
nucleation time is longer than 100 ns, glass configurations can be
generated in simulations by cooling the liquid both at low and at high
pressure.  We shall exploit this fact in the present study.

We simulate a system of 216 molecules, using the simple
point charge extended (SPC/E) model of water \cite{berensen}. This model
has been studied extensively, and the $\rho$ and $T$ dependence of
structural and dynamic properties in equilibrium are known.  In
particular, the SPC/E model reproduces the thermodynamic anomalies of
water---e.g., it produces a maximum in $\rho$ \cite{pooleSPC/E}. At
$\rho=0.94$ g/cm$^3$ at low $T$, the model describes well the LDA
structure.  We integrate the equation of motion using a time step of
$1$~fs and implement the reaction field method to account for long range
forces. Results are averaged over 16 independent realizations.  LDA ice
configurations are generated by cooling to 77 K equilibrium liquid
configurations at $\rho=0.90$~g/cm$^3$.  To generate HDA configurations,
we compress LDA at $T=77$~K to $P>1$~GPa at a rate of $5
\times10^{-5}$~g/cm$^3$/ps \cite{ourPRL}. The HDA to VHDA transition is
studied heating HDA at constant pressure from $T=77$~K using a heating
rate of $30$~K/ns.  We have used the same heating/cooling and
compression/decompression rates in all our calculations.

\begin{figure}
\centerline {
\includegraphics[width=3in]{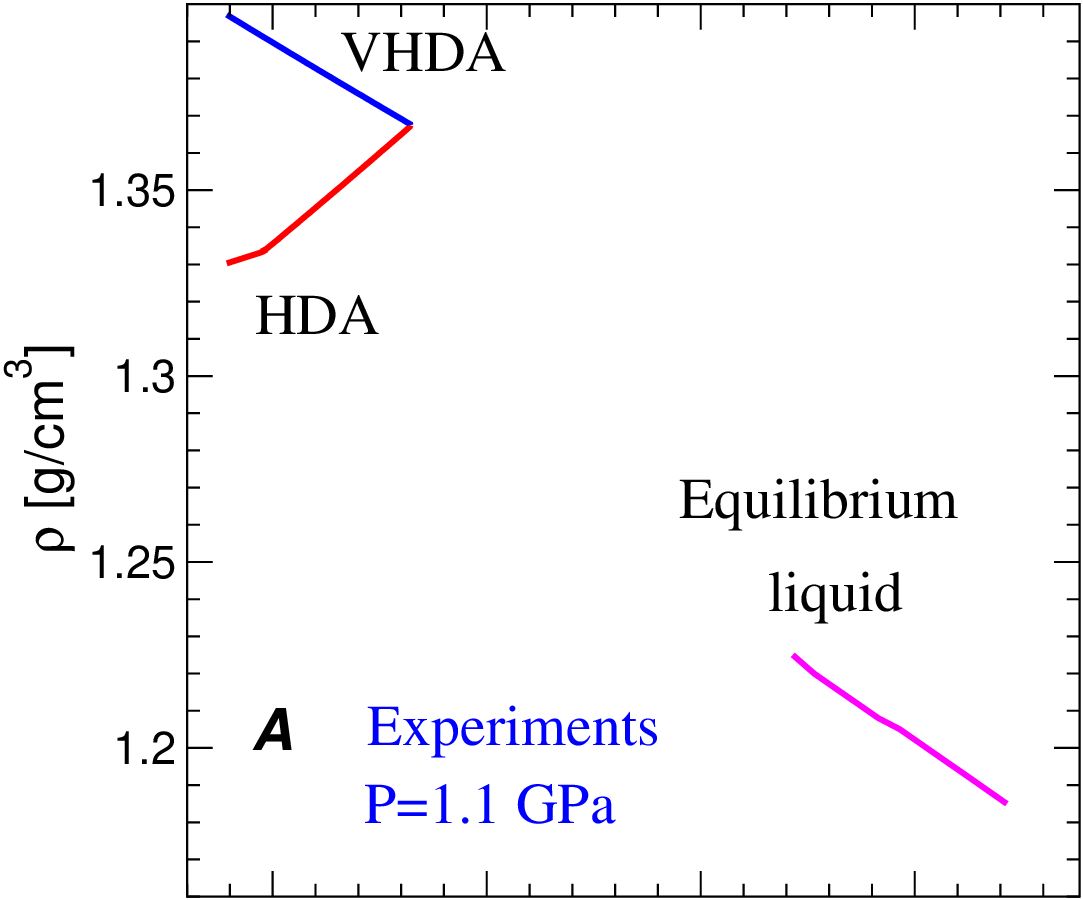}
}
\centerline {
\includegraphics[width=3in]{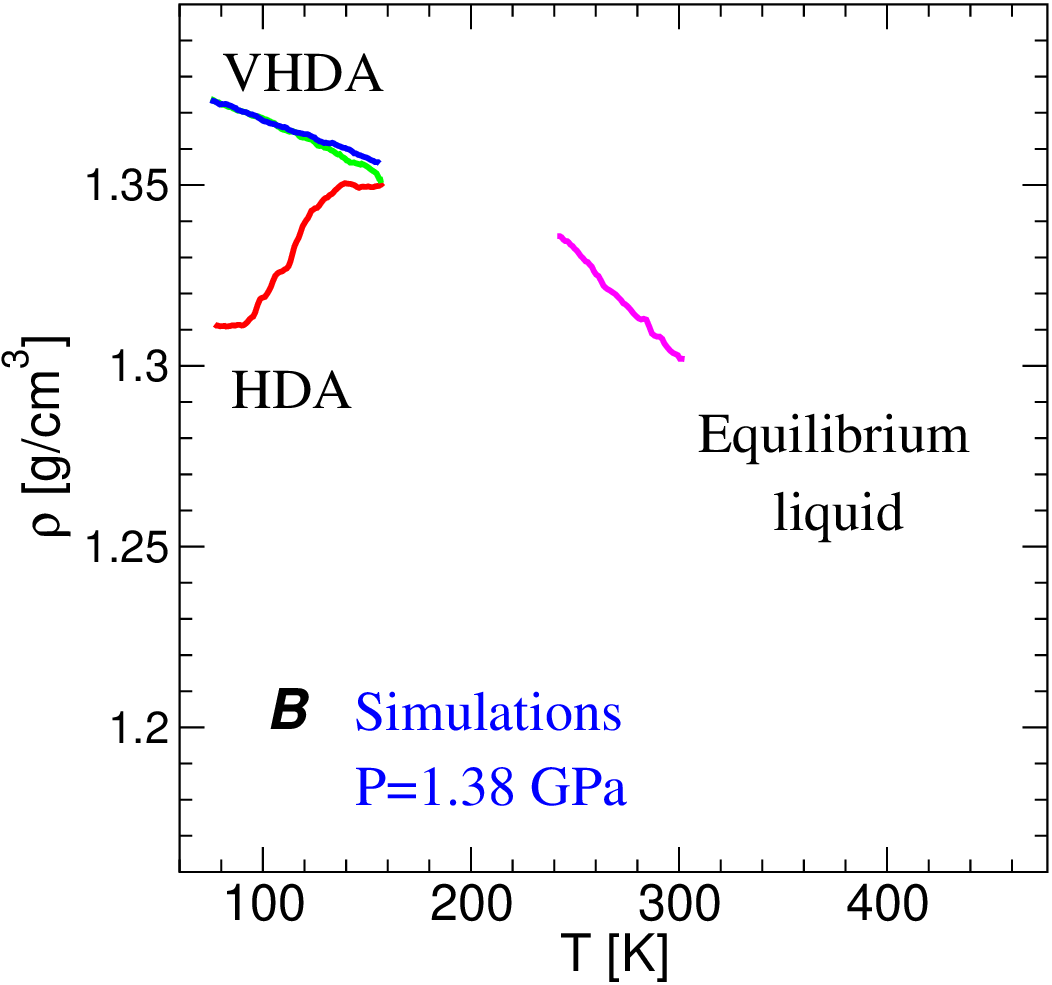}
}\vspace{0.3cm}
\caption{ Experimental data (from Ref. \cite{loerting}) and simulation
  results for the temperature dependence of the density $\rho$ during
  the conversion of HDA to VHDA at high pressure (red line) and for the
  cycling of VHDA between 165 and 77 K (green and blue lines).  The
  equilibrium $\rho$ data in the liquid state (magenta line) from the
  HGK equation of state \cite{hgk} and for SPC/E potential are also
  shown.  Numerical results for different pressure values are
  qualitatively similar to the results shown in the figure for P=1.38
  GPa. The density difference between HDA and VHDA decreases on
  increasing $P$. We also find that the density dependence of the VHDA
  recovered at T=77 K and ambient pressure varies between 1.22 and 1.28
  g/cm$^3$. }
\label{fig:sim-exp}
\end{figure}

Figure \ref{fig:sim-exp} compares, for a high pressure isobar, available
experimental data (Fig. \ref{fig:sim-exp}A) and simulation results
(Fig. \ref{fig:sim-exp}B).  In Fig. \ref{fig:sim-exp}A, the density
$\rho$ vs. temperature $T$ for equilibrium liquid water is complemented
with the $\rho$ vs. $T$ data of Ref.~\cite{loerting2} obtained by
heating HDA from 77 K up to 165 K, and then cooling back to 77 K.  VHDA
is the glass that results from densification of HDA during heating under
high pressure.  VHDA can be cycled, at constant pressure, between 77 K
and 165 K without significant further density changes, leading us to
hypothesize that VHDA is the more stable form of HDA.  Our hypothesis is
consistent with Mishima's observation that HDA samples annealed to
130-150 K at 1-1.5 GPa are characterized by identical scattering
patterns \cite{mish2}.  Relaxation of HDA to VHDA, and the
irreversibility of such transformation upon further temperature-cycling,
are facts reminiscent of slow heating of hyperquenched glasses, so the
conversion of HDA to VHDA can be interpreted as a temperature-driven
partial equilibration of the sample.

In simulations HDA is generated---in analogy with the experimental
procedure---by compressing LDA or ice I$_h$ at $77 K$ \cite{tse}, while
LDA is generated by cooling of the liquid at ambient pressure.  Figure
\ref{fig:sim-exp}B shows $\rho$ as HDA is heated.  At about $100$~K,
$\rho$ begins to increase, reaching at around $140$~K a value $0.04$
g/cm$^3$ larger than the density of HDA at $77$~K.  Decreasing $T$ back
to $77$~K does not regenerate the HDA density, but instead $\rho$
increases, in agreement with experimental results.  The resulting denser
material---which we identify with VHDA---can be cycled back and forth
without significant changes in $\rho$.

\begin{figure}
\centering {
\includegraphics[width=3in]{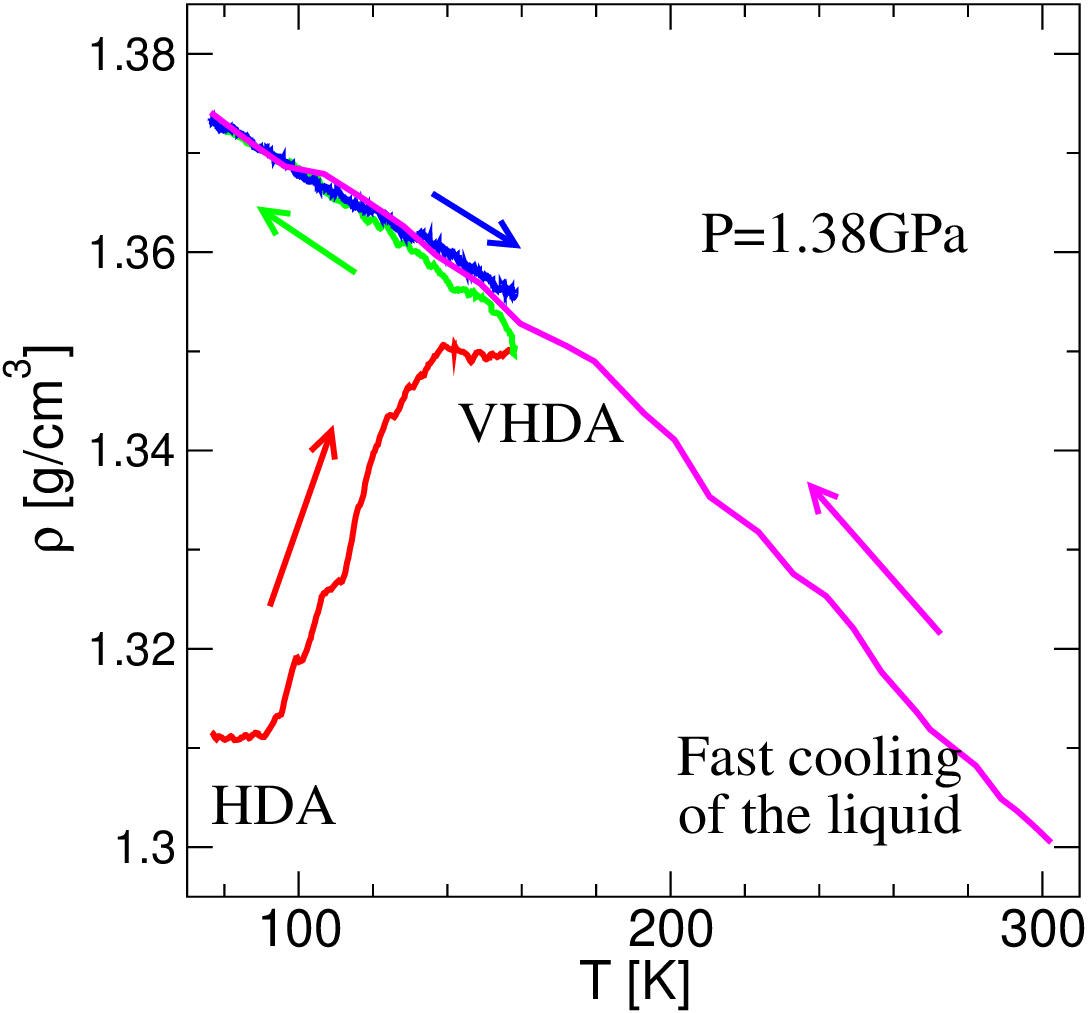}
}\vspace{0.3cm}
\caption{Comparison between the temperature-dependence of the density in
  VHDA (from Fig.\protect\ref{fig:sim-exp}B) and the
  temperature-dependence of the density during continuous cooling of the
  liquid down to 77 K, at a quenching rate of $-10^4$~K/ns.}
\label{fig:comparison}
\end{figure}

\begin{figure}
\centering {
\includegraphics[width=3in]{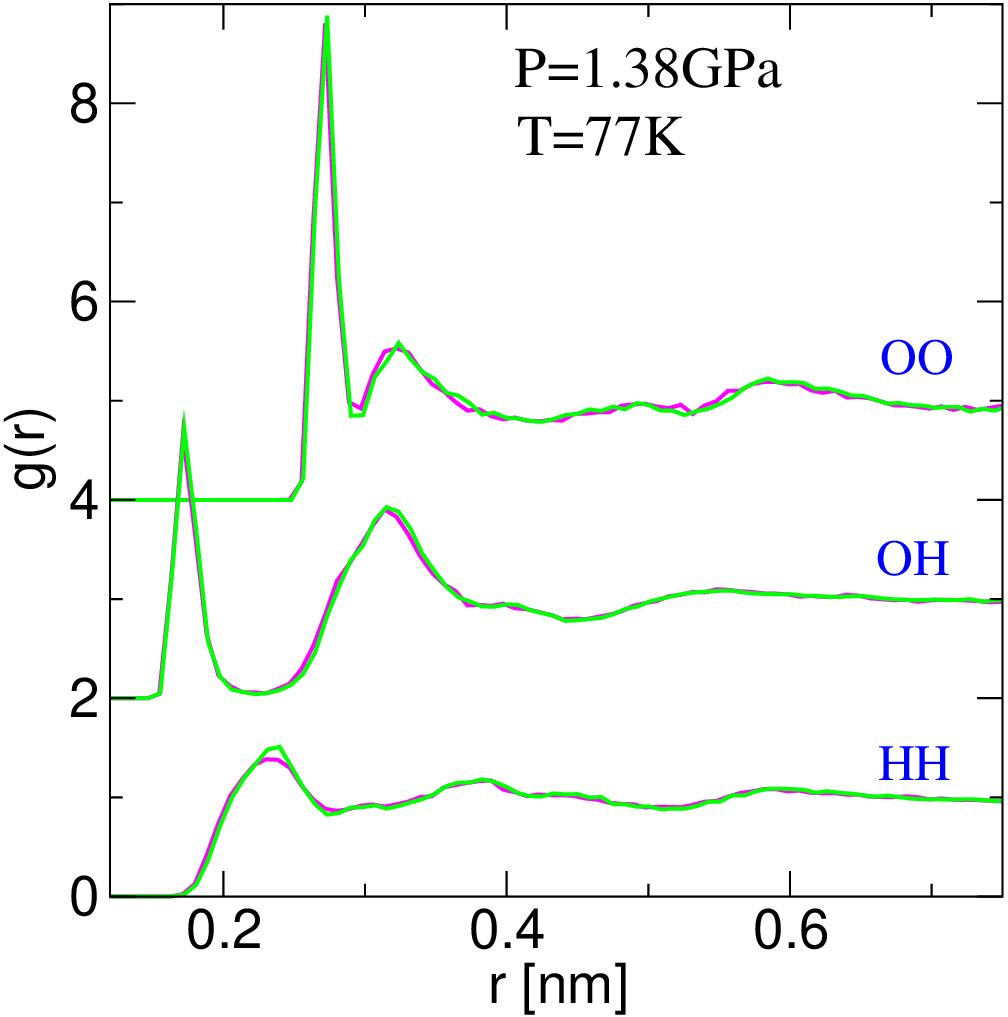}
}
\vspace{0.3cm}
\caption{Comparison between the structural properties of VHDA (green)
  and the glass obtained by cooling the liquid under pressure calculated
  at $T=77$ K and $P=1.38$ GPa (magenta).  Both oxygen-oxygen,
  oxygen-hydrogen, hydrogen-hydrogen radial distribution functions are
  shown. Despite the extremely different previous histories, the VHDA
  structure appears identical to the structure of the liquid cooled
  under pressure.}
\label{fig:gr}
\end{figure}

Next we test the hypothesis that VHDA is the glass that would be
generated by cooling the high pressure liquid.  To this end,
Fig.~\ref{fig:comparison} compares $\rho$ for isobaric cooling of the
liquid with $\rho$ for VHDA, two systems generated by {\it completely
different thermal and pressure histories}.  The $\rho$ data display
remarkable similarity, supporting the possibility that VHDA is indeed
the physical realization of the glass generated by quenching the high
pressure liquid.  To further support this interpretation,
Fig.~\ref{fig:gr} shows that the radial distribution functions of VHDA
and of the glass obtained by isobaric cooling of the liquid are indeed
indistinguishable.
 
The results presented in Figs.~\ref{fig:comparison} and \ref{fig:gr}
indicate that compression of LDA at 77 K generates a system that is
kinetically trapped due to the low temperature.  This system relaxes to
the more stable VHDA, whose structure is identical to the glass
generated by quenching the high pressure supercooled liquid.  In this
respect, not HDA and LDA but rather VHDA and LDA should be thought of as
the two distinct glassy states arising from the two distinct liquids
associated with the hypothesized line of liquid-liquid phase
transitions.
 
\begin{figure}
\centering {
\includegraphics[width=3in]{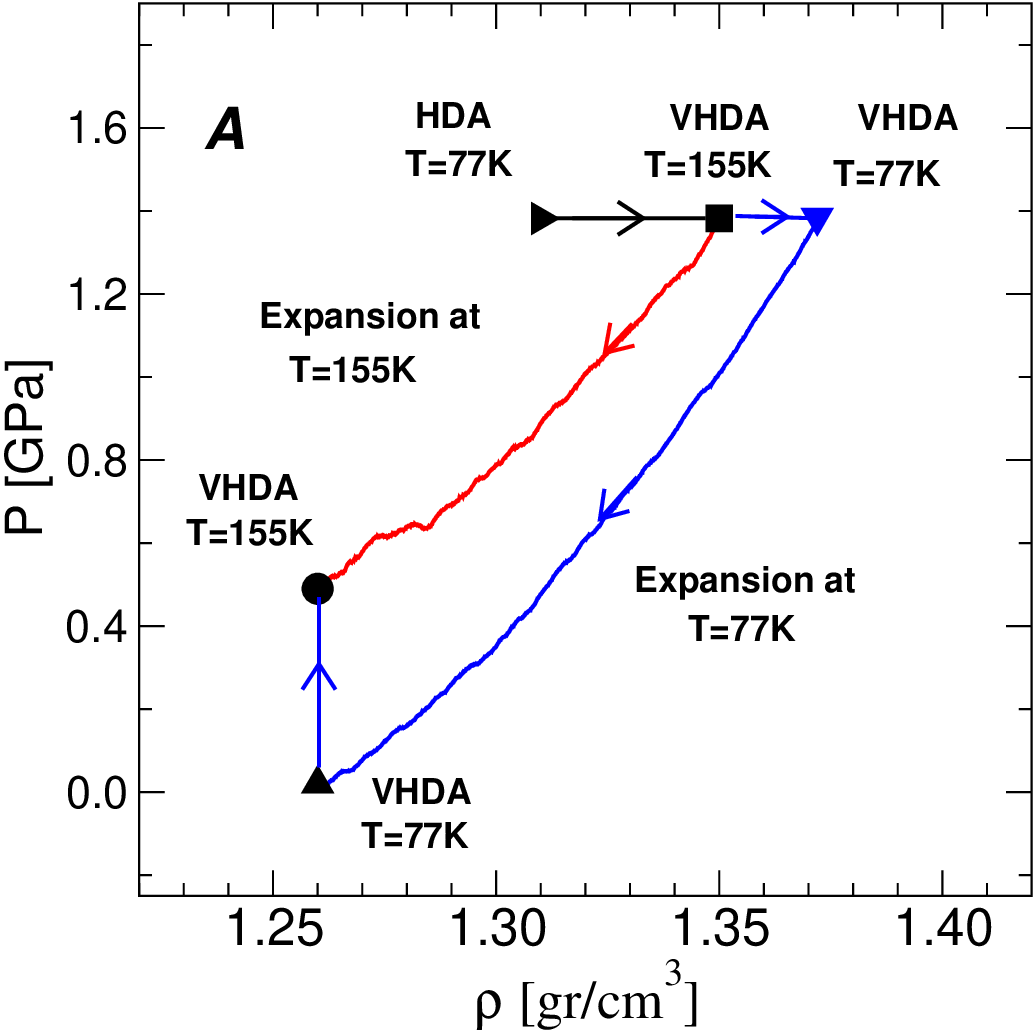}
}
\vspace{1cm}
\centering {
\includegraphics[width=3in]{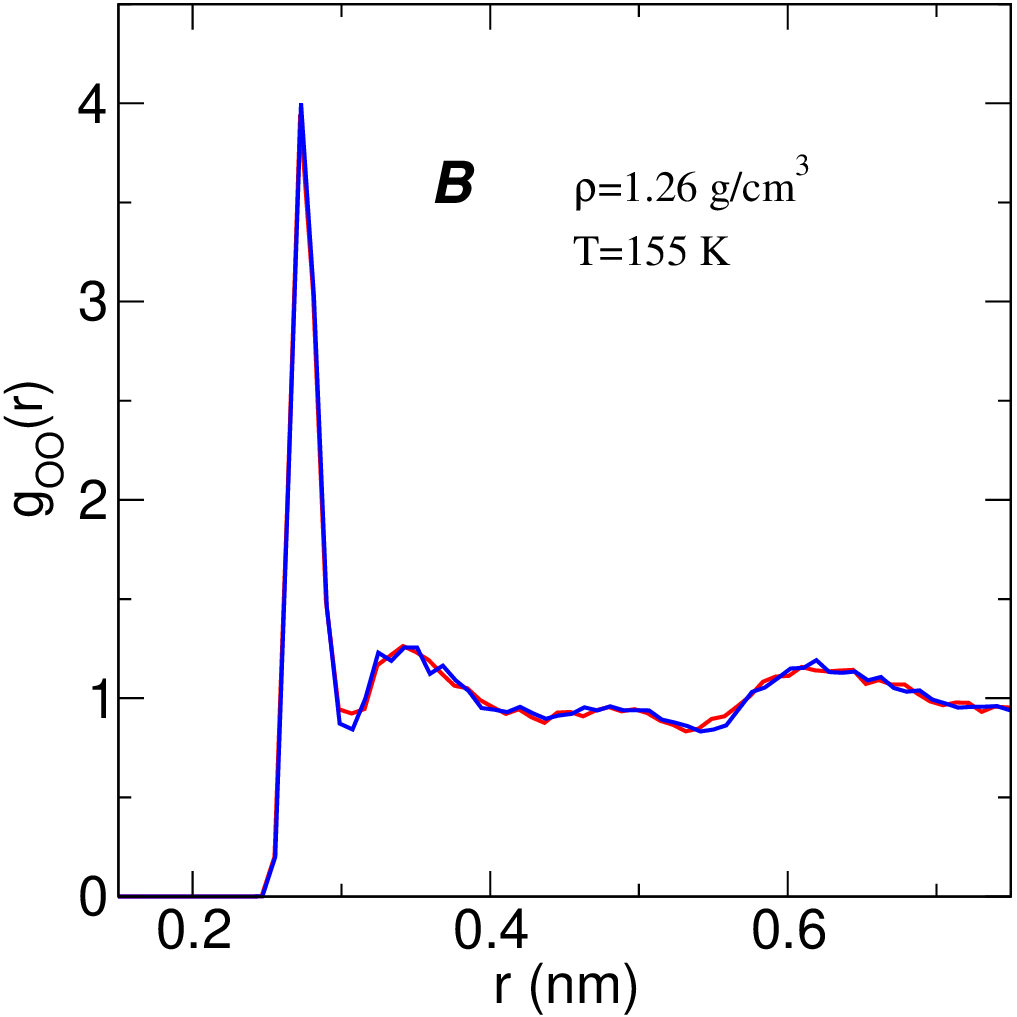}
}
\vspace{0.3cm}
\caption{Demonstration that at $T=155$K and $\rho=1.26$ g/cm$^3$ two
  glasses, each with completely different histories (red and blue
  curves), are identical both (A) thermodynamically and (B)
  structurally.  (A) $P-\rho$ diagram. Blue curve: recovering of the
  VHDA at ambient pressure at $T=77 K$ followed by an isochoric heating
  at density 1.26 g/cm$^3$ up to $T=155 K$. Red curve: isothermal
  decompression of VHDA at $T=155 K$ down to density 1.26 g/cm$^3$. The
  final product is independent of the path despite the different
  histories. The important point is that no transformation from VHDA to
  HDA is observed.  (B) Comparison between the radial distribution
  function of the two glasses obtained following the blue and the red
  paths in (A).}
\label{fig:cycle}
\end{figure}

If indeed HDA is a partially-equilibrated glass generated by the
compression technique, then one expects that VHDA should not convert to
HDA under any transformation which does not involve a different
``intermediate'' phase (such as LDA, or a crystal phase).  To confirm
this expectation, we recover VHDA at $T=77 K$ and ambient pressure with
$\rho=1.26$~g/cm$^3$.  Next we isochorically heat this system to $155$~K
(Fig.~\ref{fig:cycle}).  We compare the structure of the resulting
system with the structure of the VHDA glass recovered at the same
density at $T=155$ K.  We find that the two glasses, which are generated
from two completely different histories, are identical both
thermodynamically (Fig. 4A) and structurally (Fig. 4B), and further we
find that no VHDA to HDA transition takes place along the loop.  Our
conclusion motivates the need for a definitive experimental test that no
VHDA $\to$ HDA transition occurs.  When VHDA is heated isochorically
from 77 to 140 K, from a starting pressure of 0.02 GPa, the width of the
x-ray pattern remains narrow like VHDA \cite{loerting} (and hence does
not appear to revert to HDA); however the position of its first maximum
shifts (about 2 degrees) in the direction of HDA (cf. Figs. 1A', 1B',
and 1D' of Ref.\cite{loerting}).

In summary, our simulations suggest that VHDA is not a new form of glassy 
water but it is the result of annealing HDA upon heating at high pressure. 
We also find that VHDA is the glass obtained by fast quenching of high
pressure water, i.e. HDL. How HDA and VHDA behave upon decompression to
normal pressure and the phase diagram of amophous water are relevant topics
that will be addressed in a separate work.

We thank NSF, MIUR Cofin 2002 and Firb
and INFM Pra GenFdt for support and the Boston University Computation
Center for a generous allocation of CPU time.

\newpage 

\end{document}